\documentclass[a4paper,11pt]{article} \pdfoutput=1

\usepackage{jcappub}
\usepackage[T1]{fontenc}
\usepackage{physics}
\usepackage{bm}
\usepackage{subcaption}
\usepackage{fontawesome}
\newcommand{\rhn}{N}
\newcommand{\dm}{\mathrm{DM}}

\newcommand{\cL}{\mathcal{L}}

\newcommand{\cK}{\mathcal{K}}
\newcommand{\cJ}{\mathcal{J}}
\newcommand{\cS}{\mathcal{S}}

\newcommand{\ChiPT}{\mathrm{ChiPT}}

\newcommand{\cw}{{c_{W}}}
\newcommand{\sw}{{s_{W}}}
\newcommand{\fpi}{{f_{\pi}}}
\newcommand{\gf}{{G_{F}}}

\newcommand{\vudc}{{V^{*}_{ud}}}
\newcommand{\vusc}{{V^{*}_{us}}}

\newcommand{\piz}{{\pi^{0}}}
\newcommand{\pip}{{\pi^{+}}}
\newcommand{\pim}{{\pi^{-}}}
\newcommand{\kp}{{K^{+}}}
\newcommand{\km}{{K^{-}}}
\newcommand{\kz}{{K^{0}}}
\newcommand{\kzb}{{\bar{K}^{0}}}

\newcommand{\sibar}{{\bar{\sigma}}}

\newcommand{\Hazma}{{\itshape\bfseries Hazma}}
\newcommand{\Herwig}{{\itshape\bfseries Herwig}}
\newcommand{\Pythia}{{\itshape\bfseries Pythia}}
\newcommand{\PPPC}{{\itshape\bfseries PPPC4DMID}}
\newcommand{\HDMSpectra}{{\slshape\bfseries HDMSPectra}}

\newcommand{\GitHubLink}[1]{{\href{https://github.com/LoganAMorrison/blackthorn}{#1}}}

\title{\begin{center}\boldmath Sterile Neutrinos from Dark Matter: \\ A {$\nu$} Nightmare? \end{center}}

\author[a,b]{Logan Morrison,}
\author[a,b]{Stefano Profumo,}
\author[c]{Bibhushan Shakya}

\affiliation[a]{Department of Physics, 1156 High St., University of California Santa Cruz, Santa Cruz, CA 95064, USA}
\affiliation[b]{Santa Cruz Institute for Particle Physics, 1156 High St., Santa Cruz, CA 95064, USA}
\affiliation[c]{Deutsches Elektronen-Synchrotron DESY, Notkestr.~85, 22607 Hamburg, Germany}

\emailAdd{loanmorr@ucsc.edu}
\emailAdd{profumo@ucsc.edu}
\emailAdd{bibhushan.shakya@desy.de}
 
\subheader{DESY-22-176}

\abstract{We provide a comprehensive study of observable spectra from dark matter pair-annihilation or decay into sterile (right-handed) neutrinos. This occurs, for instance, in neutrino portal dark matter models, where a sterile neutrino acts as the portal between dark matter and the Standard Model sector. The subsequent decays of right-handed neutrinos produce detectable Standard Model particles, notably photons, positrons, and neutrinos. We study the phenomenology of models where the right-handed neutrino masses are below the GeV scale, as well as models where they are at, or significantly heavier than, the TeV scale. In both instances, and for different reasons, the standard tools, including Monte Carlo simulations, are both inadequate and inaccurate. We present the complete framework to compute the relevant branching ratios for right-handed neutrino decays and the spectra of secondary photons, positrons, and neutrinos for a broad range of dark matter and right-handed neutrino masses. We discuss the general features of such signals, and compare the spectra to standard signals from dark matter annihilation/decay into bottom quarks. Additionally, we provide open source code\footnote{The code is available at \href{https://github.com/LoganAMorrison/blackthorn}{https://github.com/LoganAMorrison/blackthorn}.} that can be used to compute such spectra.\href{https://github.com/LoganAMorrison/blackthorn}{\faGithub}}

\begin{document}
\maketitle
\flushbottom

\section{Introduction}\label{sec:intro}
The search for the particle nature of the cosmological dark matter (DM) continues without any conclusive non-gravitational signals so far (for a review, see \cite{rpp}). In particular, the search for the annihilation or decay products from DM in photons, neutrinos, or charged cosmic rays, while yielding null results, continue to grow more sensitive with the advent of new experimental programs. Such searches are often cast in a model-independent manner, assuming that a single final state dominates the annihilation or decay process. This final state particle is generally taken to be a Standard Model (SM) constituent, such as a fermion or gauge boson; however, this need not be the case in several realistic dark matter frameworks,  and the annihilation or decay products could instead be made up of particles beyond the Standard Model (BSM). In this paper, we study a particular instance of this that is extremely well motivated: dark matter annihilation or decay into sterile, right-handed (i.e. not charged under SU(2)) neutrinos. Right-handed neutrinos (RHN), also dubbed sterile neutrinos or heavy neutral leptons (HNL), constitute one of the best-motivated extensions of the Standard Model (SM), featuring in many models of neutrino mass generation. In such models,
RHNs are often part of an extended sector that also contains dark matter. Such
frameworks have been extensively studied in the literature under the broad
umbrella of {\em neutrino portal dark matter} (see e.g. Ref.\,\cite{Falkowski:2009yz,Macias:2015cna,Escudero:2016ksa,Escudero:2016tzx,Tang:2016sib,Batell:2017cmf,Batell:2017rol,Shakya:2018qzg,Patel:2019zky}),
where the RHNs act as the portal connecting DM to the visible sector.

Since RHNs do not participate in any of the SM interactions, DM annihilation or decay entirely into RHNs could give rise to a ``nightmare scenario'' for indirect dark matter detection, where the signals arising from DM annihilation/decay are extremely suppressed compared to standard signals with SM final states, or even entirely invisible. The existence of mixing between the sterile and active (SM) neutrinos, however, opens up the possibility that lower energy SM products of RHN decay could still allow for detectable indirect detection signals. The goal of this study is to explore in detail the indirect detection signals arising from DM annihilation/decay into RHNs across a broad range of masses, and compare these with standard signals from decay into SM states in order to establish whether this could indeed constitute a ``nightmare scenario'' for indirect detection.


If dark matter is lighter than the RHN, it annihilates or decays directly to
SM states via off-shell RHNs, via the mixing between sterile and active neutrinos (see e.g.\,\cite{Falkowski:2009yz,Patel:2019zky}), albeit with rates suppressed by this mixing angle.   On the other hand, if DM is heavier, it tends to annihilate or decay {\em exclusively} to RHNs, and subsequent decays of the RHNs into SM particles then give rise to visible
signals.\,\footnote{Such signals have been employed in the past to explain various putative dark matter signals such as the Galactic Center excess
\cite{Tang:2015coo} and high energy neutrinos at IceCube \cite{Roland:2015yoa}.} In this paper, we will focus on the latter scenario, as the observable spectra of SM states in this case is expected to differ significantly from the ``standard'' SM final state spectra generally considered in the literature, where DM is assumed to annihilate/decay directly into SM states. Such signals are also fairly insensitive to the exact nature of the underlying
model, since DM annihilation (or decay) produces an isotropic
distribution of RHNs with energy $m_{DM}(\text{or }m_{DM}/2)$. While the decay widths of the RHNs are suppressed by their small mixing with SM states, such decays can generally
be considered prompt on astrophysical scales for the purposes of indirect detection, and hence independent of the size of the mixing angle (however, exceptional cases can occur for DM annihilation/decay in the sun
\cite{Allahverdi:2016fvl}, or RHNs with extremely long lifetimes
\cite{Gori:2018lem}). The spectra of visible particles are thus essentially determined by only two
parameters: the dark matter mass $m_{DM}$ and the RHN mass
$m_{N}$. 

Indirect detection of DM
annihilation into RHNs has only been studied in the literature for very specific cases:
for $m_N=1-5$ GeV in \cite{Allahverdi:2016fvl}, and for $m_N=10-1000$ GeV in
\cite{Campos:2017odj}. A primary goal of this paper is to perform an extensive study of such signals in terms of these parameters, highlighting the main qualitative features of observable signals of DM annihilating or decaying into RHNs over a broad range of mass scales. At present, tools designed to compute the decay products of sterile neutrinos are essentially limited to where the yields of simulation tools such as \Pythia~\cite{bierlich2022comprehensive} or {\itshape\bfseries HERWIG}~\cite{Bellm:2015jjp} are reliable -- above approximately 5 GeV and below the TeV scale.  Below the few GeV scale, RHN decay involving strongly interacting particles proceeds directly to hadronic states, rather than jets that subsequently hadronize and fragment, as assumed in the above-mentioned codes. In the multi-TeV energy range, the bremstsrahlung of electroweak bosons becomes important, and, again, is not included in existing codes but requires the use of specialized tools. In this paper, we provide results and computational tools to overcome these shortcomings at both low and high energy scales, thereby enabling the computation of observable spectra in these regimes. We provide open source code that can be used to compute such spectra for a wide range of dark matter and RHN masses on 
\href{https://github.com/LoganAMorrison/blackthorn}{GitHub}\cite{BlackthornGitHub}.

This paper is organized as follows. Section~\ref{sec:framework} describes the field-theoretical framework of our study. Section~\ref{sec:formalism} covers the details of the calculation of visible spectra and spectra from $n$-body decays of RHNs. Section~\ref{sec:results} contains the main results of our paper: the decay branching ratios of RHNs into various SM final states across a broad range of RHN masses, the spectra of photons, positrons, and neutrinos from DM annihilation/decay into RHNs, and comparisons with analogous spectra from DM annihilation/decay into $b\bar{b}$. The main findings of the paper are summarized in Section~\ref{sec:conclusions}.

\section{Framework}
\label{sec:framework}


We denote the dark matter particle as $\chi$, and the right-handed neutrino as $N$. Our studies cover two classes of DM models: one in which DM {\em decays} into two RHNs (\(\chi
\to \rhn\rhn\)), and another where DM {\em annihilates} into a RHN pair (\(\chi\bar{\chi} \to \rhn\rhn\)).  
For the purposes of this paper, the exact forms of the interactions
are irrelevant: the spectra we
 present and study do not depend on the specifics of the interactions, since the
energy of the RHNs (which in turn dictates their decay spectra) only depends on the center-of-mass energy of the process, which is
fixed by the dark matter mass and velocity. Thus, the visible spectra are primarily determined by the details of the RHN interactions with SM fields.

For simplicity, we assume the existence of a single Majorana RHN that couples to the SM via a
Yukawa interaction. In two-component spinor notation, the terms in the
Lagrangian density containing the neutrinos are
\begin{align}
	\mathcal{L}
	 & \supset
	i\hat{\rhn}^{\dagger}\bar{\sigma}_{\mu}\partial^{\mu}\hat{\rhn}
	- \frac{1}{2}\hat{m}_{\hat{\rhn}}\qty(\hat{\rhn}\hat{\rhn} + \hat{\rhn}^{\dagger}\hat{\rhn}^{\dagger})
	+ \epsilon^{ab}Y^{i}_{\nu}\Phi_{a}L_{bi}\hat{\rhn}
	.
\end{align}
Here \(\epsilon^{ab}\) is the two-dimensional Levi-Civita tensor, with \(\epsilon^{12}=+1\), \(\Phi_{a}\) is the Higgs SU(2) doublet, \(L_{bi}\) is the lepton doublet for
the \(i\)th generation (assumed to be such that the charged lepton mass matrix is diagonal),
and \(\hat{\rhn}\) is the RHN represented as a two-component
Majorana spinor. The vector \(Y^{i}_{\nu}\) is a Yukawa vector coupling the
\(i\)th lepton doublet to the RHN. Expanding the Higgs around its vacuum
expectation value, the neutrino mass terms are
\begin{align}
	\mathcal{L}_{\mathrm{mass},\nu}
	                        & \supset
	- \frac{1}{2}\hat{m}_{\hat{\rhn}}\qty(\hat{\rhn}\hat{\rhn} + \hat{\rhn}^{\dagger}\hat{\rhn}^{\dagger})
	- \frac{v}{\sqrt{2}}Y^{i}_{\nu}\hat{\nu}_{i}\hat{\rhn}
	=
	-\frac{1}{2}
	\mathcal{N}^{T}
	\mqty(\bm{0}_{3\times3} & \frac{v}{\sqrt{2}}Y_{\nu} \\ \frac{v}{\sqrt{2}}Y^{T}_{\nu} & \hat{m}_{\hat{\rhn}})
	\mathcal{N},
\end{align}
where \(\mathcal{N} = \mqty(\hat{\nu}_1 & \hat{\nu}_3 &\hat{\nu}_3&\hat{\rhn})^{T}\)
is a vector composed of all neutrinos. For simplicity, we will assume that only
a single entry of \(Y_{\nu}\) is non-zero, i.e. we set \(Y^{k}_{\hat{\nu}} = y\)
and \(Y^{i}_{\hat{\nu}} = 0\) for \(i\neq k\), which corresponds to $N$ mixing with only a single SM neutrino flavor. In this case, we may safely drop
the active neutrinos \(\hat{\nu}_{i}\) for \(i\neq k\) from the mass matrix and
take them to be mass eigenstates. Then, the neutrino mass terms reduce to
\begin{align}
	\mathcal{L}_{\mathrm{mass},\nu}
	                    & \supset
	-\frac{1}{2}
	\mqty(\hat{\nu}_{k} & \hat{\rhn})
	\underbrace{\mqty(0 & \frac{v}{\sqrt{2}}y \\ \frac{v}{\sqrt{2}}y & \hat{m}_{\hat{\rhn}})}_{M_{\nu}}
	\mqty(\hat{\nu}_{k}                       \\ \hat{\rhn}).
\end{align}
The neutrino mass matrix can be diagonalized using the Takagi diagonalization method via a
unitary matrix \(\Omega\), where
\(\Omega^{T}M_{\nu}\Omega = \mathrm{diag}(m^{k}_{\nu}, m_{\rhn})\). The
explicit form of \(\Omega\) is
\begin{align}
	\Omega = \mqty(-i\cos\theta & \sin\theta \\ i\sin\theta & \cos\theta),
\end{align}
which can easily be checked to be unitary. The parameters
\(y, \hat{m}_{\hat{\rhn}}\) can be translated to \(m_{\rhn}\) and \(\theta\)
as:
\begin{align}
	y & = \frac{\sqrt{2}m_{\rhn}\tan\theta}{v}, & \hat{m}_{\hat{\rhn}} & = m_{\rhn}\qty(1-\tan^2\theta).
\end{align}
In addition, the left-handed neutrino mass is \(m^{k}_{\nu} = m_{\rhn}\tan^2\theta\).
To obtain the interactions between the RHN and SM particles, we use
\begin{align}\label{eqn:framework:gauge_to_mass_eigenstates}
	\hat{\nu}  & = -i\cos\theta\nu_{k} + \sin\theta\rhn, &
	\hat{\rhn} & = i\sin\theta\nu_{k} + \cos\theta\rhn,
\end{align}
where the unhatted fields \(\nu_{k}\) and \(\rhn\) are mass eigenstates.

Making the replacements given in Eqn.\,(\ref{eqn:framework:gauge_to_mass_eigenstates}), the following interaction Lagrangian emerges for $N$ and \(\nu_{k}\):
\begin{align}\label{eqn:framework:ew_interaction_lagrangian}
	\cL_{\mathrm{int},\nu} & =
	\cL_{\nu W^{\pm}}
	+ \cL_{\nu Z}
	+ \cL_{\nu h}
	+ \cL_{\nu G^{\pm}}
	+ \cL_{\nu G^{0}},
\end{align}
where the \(W\), \(Z\) and Higgs Lagrangians containing the relevant interaction terms are
\begin{align}
	\cL_{\nu W^{\pm}}
	 & =
	\frac{e}{\sqrt{2}\sw}\qty(-i\cos\theta W^{-}_{\mu}\ell^{\dagger}_{k}\bar{\sigma}_{\mu}\nu_{k}
	+ \sin\theta W^{-}_{\mu}\ell^{\dagger}_{k}\bar{\sigma}_{\mu}\rhn) + \mathrm{h.c}; \\
	\cL_{\nu Z}
	 & =
	\frac{e}{2\cw\sw}Z_{\mu}\qty[
		\cos^2\theta\nu^{\dagger}_{k}\bar{\sigma}_{\mu}\nu_{k}
		+ \sin^2\theta\rhn^{\dagger}\bar{\sigma}_{\mu}\rhn
		+ \qty(i\cos\theta\sin\theta \nu_{k}^{\dagger}\bar{\sigma}_{\mu}\rhn + \mathrm{c.c})
	];                                                                                \\
	\cL_{\nu h}
	 & =
	-\frac{h\sin\theta}{v}\qty[
		m_{\nu}\cos\theta\nu_{k}\nu_{k}
		+ m_{\rhn}\sin\theta\rhn\rhn
		+ \qty(i m_{\rhn}\cos\theta\rhn\nu_{k} + \mathrm{c.c.})
	] + \mathrm{c.c.};
\end{align}
and the Goldstone Lagrangians are
\begin{align}
	\cL_{\nu G^{\pm}}
	 & =
	\frac{\sqrt{2}G^{+}}{v}\ell_{k}\qty(
	im_{\nu}\sin\theta\nu_{k}
	+ m_{\rhn}\sin\theta\rhn)
	+ \frac{\sqrt{2}m_{\ell_{k}}G^{-}}{v}\bar{\ell}_{k}\qty(
	i\cos\theta\nu_{k}
	+ \sin\theta\rhn)
	+ \mathrm{c.c.};
	\\
	\cL_{\nu G^{0}}
	 & =
	-\frac{i G^{0}\sin\theta}{v}\qty[
		m_{\nu}\cos\theta\nu_{k}\nu_{k}
		+ m_{\rhn}\sin\theta\rhn\rhn
		+ \qty(i m_{\rhn}\cos\theta\rhn\nu_{k} + \mathrm{c.c.})
	] + \mathrm{c.c.}.
\end{align}
Here \(c_{W},s_{W}\) are the cosine and sine of the weak
mixing angle, \(e\) is the electromagnetic gauge coupling, and ``\(\mathrm{c.c.}\)'' stands
for ``complex conjugate''.

The Lagrangian in Eqn.\,(\ref{eqn:framework:ew_interaction_lagrangian}) is appropriate
for $m_N \gtrsim\mathcal{O}$(GeV). For sub-GeV masses, we
need to integrate out the \(W\), \(Z\) and Higgs bosons and match onto an effective Lagrangian that takes into account QCD bound states. For this purpose, we use the Chiral Lagrangian \cite{GASSER1985465,Scherer2003,WEINBERG1979327}. We match onto the chiral Lagrangian in two steps: first we integrate out the heavy bosons, then we match the resulting Lagrangian onto the chiral Lagrangian. We discuss the details of this procedure in Appendix \ref{app:chiral} (see also Ref.~\cite{Coogan:2021sjs}).

\section{Formalism}\label{sec:formalism}

In this section, we discuss the details of the formalism we employ for our spectra calculations. We will study the spectra of gamma rays, positrons, and neutrinos, but ignore antiprotons, since observational prospects for antiprotons are significantly weaker than for the other states for neutrino portal models. 

We first discuss the computational tools we use in our study, before turning to the details of the calculation of the spectra of visible particles, and instrumental effects.


\subsection{Computational Tools}

Depending on the RHN mass, different tools are needed to generate
the RHN decay spectra. We partition the RHN mass range into three regions: high-mass (\(1 \ \mathrm{TeV} \lesssim m_{\rhn} \lesssim M_{\mathrm{pl}}\)), 
EW (\(5 \ \mathrm{GeV} \lesssim m_{\rhn}\lesssim 1 \ \mathrm{TeV}\)),
and sub-GeV (\(m_{\rhn}\lesssim 500 \ \mathrm{MeV}\)). The intermediate regime between 0.5 GeV and 5 GeV has only recently been addressed with a new release of the \Hazma{}~\cite{hazmaMeetsHerwig} code, albeit only for vector-mediator dark matter models: the key issue is the necessity to include a large number of hadronic final states, weighed by form factors whose calculation lies deep in the non-perturbative QCD regime (the new \Hazma{} release utilizes the vector meson dominance scheme in conjunction with $e^+e^-$ data). Due to such complications, we omit RHN masses between 0.5 GeV and 5 GeV in this work. In each of the three regions we consider, we use a different software package designed
specifically to generate spectra in the given mass range, as described below.

\paragraph*{\(\bm{1 \ \mathrm{TeV} \lesssim m_{\rhn} \lesssim M_{\mathrm{pl}}}\):}
For heavy $m_N$, we use \HDMSpectra{}~\cite{bauer2021dark}. \HDMSpectra{} computes
spectra of stables particles using pre-computed fragmentation functions \(D^{b}_{a}(x;\mu_{Q},\mu_{0})\),
which describe the probability of an initial particle \(a\) at an energy scale
\(\mu_{Q}\) eventually producing a particle \(b\) at a scale \(\mu_{0}\) with
momentum \(p_{b} = xp_{a}\). The fragmentation functions are computed by
evolving them from a high energy scale \(\mu_{Q}\) (set by the mass of the
decaying particle in question) down to the electroweak scale
\(\mu_{\mathrm{EW}}\), using all SM interactions in the unbroken
\(\mathrm{SU}_{c}(3)\times\mathrm{SU}_{L}(2)\times\mathrm{U}_{Y}(1)\) theory and
partially including soft-coherence effects. During this stage, all states are approximated to be massless. The evolution is carried out using the
DGLAP equations.  At the electroweak scale, the top quark and \(W, Z\) bosons are decayed
using analytical results, and the Higgs is decayed using \Pythia{}~\cite{sjostrand2015introduction,sjostrand2008brief}. Below the
electroweak scale, \Pythia{} is used without FSR for the remaining
evolution, and FSR is handled using the Altarelli-Parisi splitting functions. The results from \HDMSpectra{} are valid for decays/annihilations with
center-of-mass energy \(\gtrsim 1 \ \mathrm{TeV}\). Above \(1 \ \mathrm{TeV}\),
the authors report an estimated accuracy of
\(\order{10\%}\) at an EeV and \(\order{20\%}\) at the Plank scale for \(x\in[10^{-3},1]\). 
When applied to our setup, this approach misses contributions from electroweak corrections to annihilations and decays of DM into sterile neutrinos; however, these corrections are suppressed by the small active-sterile mixing angle and thus should be negligible compared to the pure SM effects.

\paragraph*{\(\bm{5 \ \mathrm{GeV} \lesssim m_{\rhn}\lesssim 1 \ \mathrm{TeV}}\):}
This mass window is within the regime of validity of \Pythia{}, and we calculate the spectra using \PPPC{}~\cite{cirelli2011pppc}. \PPPC{} provides
pre-computed tables of spectra of stable particles from DM annihilations. The
tables are generated in similar fashion to \HDMSpectra{}.
\PPPC{} includes finite masses and considers only leading order EW
corrections. Here, we utilize the tables provided in \PPPC{} to generate spectra for individual final state particles resulting from sterile neutrino decays. We handle final state particle distributions and branching fractions as described below. 

\paragraph*{\(\bm{m_{\rhn}\lesssim 500 \ \mathrm{MeV}}\):}
Below the GeV scale, we use spectra for \(\mu^{\pm},\pi^{0},\pi^{\pm}\) and
\(K^{\pm}\), computed using \Hazma{}~\cite{coogan2020hazma}. \Hazma{} uses analytic 
results for the decay spectra of \(\mu^{\pm}\to e^{\pm}\nu_{e}\nu_{\mu}\) and
\(\pi^{0}\to\gamma\gamma\). The \(\pi^{\pm}\) spectra are generated by boosting
the \(\mu^{\pm}\) spectra from \(\pi^{\pm}\to\mu^{\pm}\nu_{\mu}\) and including
the radiative decay spectra from \(\pi^{\pm}\to\ell^{\pm}\nu\gamma\). For the
\(K^{\pm}\) spectra, \Hazma{} bootstraps the spectrum from the
\(K^{\pm}\to\mu^{\pm}\nu_{\mu}\), \(\pi^{\pm}\pi^{0}\), \(\pi^{\pm}\pi^{+}\pi^{-}\),
\(\pi^{\pm}\pi^{0}\pi^{0}\), \(\pi^{0}e^{\pm}\nu_{e}\), and \(\pi^{0}\mu^{\pm}\nu_{\mu}\) decay channels.

\subsection{Spectra of visible particles}

We are interested in the spectra of stable particles \(f\in\qty{\gamma,e^{+},\nu_{e},\nu_{\mu},\nu_{\tau}}\) produced from RHN decays. We broadly categorize the contributions into three groups according to means of production: from direct decay into \(f\) (\(\rhn\to f + Y\)), as secondary products from an intermediate unstable state (\(\rhn \to X + Y \to f + Y + Z \)), or from final state radiation (FSR) (\(\rhn\to X^* + Y \to X + f + Y\)), where \(X,Y,Z\) represent generic groups of final states. We thus separate the calculation of spectra into three parts:
\begin{align}
	\dv{N_{f}}{x} & = \sum_{X}\mathrm{BR}\qty(\rhn\to X)\qty[
		\eval{\dv{N_{X\to f}}{x}}_{\mathrm{direct}}+
		\eval{\dv{N_{X\to f}}{x}}_{\mathrm{decay}}+
		\eval{\dv{N_{X\to f}}{x}}_{\mathrm{FSR}}
	],
\end{align}
where \(\mathrm{BR}(\rhn\to X) = \Gamma(\rhn\to X)/\Gamma_{\rhn}\) is the
branching ratio into the final state \(X\). We define the dimensionless variable \(x\) as \(x = 2E_{f}/\sqrt{s}\), where \(\sqrt{s}\) is the center-of-mass energy ($2 m_{\dm}$ for annihilation, $m_{\dm}$ for decay).

\paragraph*{\textbf{Direct decays:}}

The differential energy spectrum in this case is given by
\begin{align}
	\eval{\dv{N_{\rhn\to f + Y}}{x}}_{\mathrm{direct}}
	 & =
	\frac{1}{\Gamma(\rhn\to f + Y)}\dv{\Gamma(\rhn\to f + Y)}{x}\,,
\end{align}
where \(\Gamma(\rhn\to f + Y)\) is the partial decay width into \(f + Y\). If
\(Y\) consists of a single state \(A\), then \(\dv*{\Gamma(\rhn\to
	f+A)}{x}=\Gamma(\rhn\to f + A)\delta(x-x_0)\), such that
\begin{align}
	\eval{\dv{N_{\rhn\to f + A}}{x}}_{\mathrm{direct}}
	 & =
	\delta(x-x_0),
	 &
	x_0
	 & =
	1 + \frac{m_{f}^2-m_{A}^2}{m_{\rhn}^{2}}.
\end{align}
If \(Y\) consists of multiple particles, the energy spectrum of \(f\)
must be computed using the full phase space. For three-body decays, i.e. \(Y = A + B\), the results is
\begin{align}
	\eval{\dv{N_{\rhn\to f + A + B}}{x}}_{\mathrm{direct}}
	 & =
	\frac{1}{\Gamma(\rhn\to f + A + B)}\frac{1}{256 m_{\rhn}\pi^3}\int_{y^{-}}^{y^{+}}\dd{y}\abs{\mathcal{M}}^{2}
\end{align}
where \(\abs{\mathcal{M}}^{2}\) is the spin-averaged squared matrix
element, and \(y = 2E_{A}/m_{N}\). The
integration limits \(y^{\pm}\) are given by
\begin{align}
	y^{\pm} & = \frac{
		\qty(2-x)\qty(\bar{x} + \mu_{A}^{2} - \mu_{B}^{2})
		\mp
		\lambda^{1/2}\qty(1,\mu_{f}^{2},\bar{x})
		\lambda^{1/2}\qty(\mu_{A}^{2}, \mu_{B}^{2}, \bar{x})
	}{2\bar{x}}\,,
\end{align}
with \(\bar{x} \equiv 1-x+\mu_{f}^{2}\), \(\mu_{\alpha} =
2m_{\alpha}/m_{\rhn}\), and \(\lambda(a,b,c)=a^{2}+b^{2}+c^{2} - 2ab -
2ac -2bc\) the K\"{a}llen-\(\lambda\) triangle function. In cases where \(Y\)
consists of 3 or more particles, the energy spectrum can be computed using
Monte-Carlo integration; however, all of the RHN decay channels of relevance (see Section \ref{sec:BRs}) consist of two or three body final states, hence the above semi-analytic expressions are sufficient for the purposes of this paper.

\paragraph*{Decays from unstable particles:}

To calculate the differential energy spectrum of \(f\) when the RHN decays into an unstable particle \(J\), which then decays
into a final state consisting of \(f\), we first compute the spectrum of \(J\),
\(\dv*{N_{J\to f}}{x}\), in the rest frame of \(J\), then boost this spectrum into
the rest frame of the RHN (see Appendix \ref{app:boost} for details).
These two steps yield a differential energy spectrum \({\dv*{N_{J\to f}}{x}}(x|E_{J})\)
conditioned on the energy of \(J\). If \(J\) has a distribution of energies
(which is the case for \(n\geq3\)), we marginalize over the energy distribution of \(J\) (computed using the methods described above).
The general spectrum from a two-step decay chain is given by
\begin{align}
	\dv{N_{\rhn\to X\to f}}{x}
	 & =
	\sum_{J \in X}
	\int \dd{E}_{J}
	{\dv{N_{J}}{E_{J}}}(E_{J})
	{\dv{N_{J\to f}}{x}}(x|E_{J})\,,
\end{align}
where \(\dv*{N_{J}}{E_{J}}\) is the energy distribution of particle \(J\). 

\paragraph*{\textbf{Final-state radiation}:}

For RHN decays into charged states, final state radiation (FSR) plays an
important role in generating accurate photon spectra. When the mass of the charged state is
small compared to the center-of-mass energy, the FSR is dominated by soft and
collinear photons. These contributions are well-approximated by the
Altarelli-Parisi splitting functions. For charged scalars and fermions, the FSR
spectrum is approximately
\begin{align}
	{\dv{N^{(\mathrm{FSR})}_{S,F\to\gamma}}{x}}(x,s)
	 & \sim
	\frac{Q_{S,F}^2\alpha_{\mathrm{EM}}}{2\pi}
	\mathcal{P}_{S,F}(x)
	\qty(\log(\frac{s(1-x)}{m_{F,S}^{2}})-1)
	 &
\text{for}~~	\frac{m_{S,F}^{2}}{s}
	 &
	\to 0.
\end{align}
Here, \(s\) is the squared center-of-mass energy and \(Q_{S,F}\) is the charge
of the scalar or fermion.  The functions \(\mathcal{P}_{S,F}(x)\) are the
scalar and fermion splitting functions
\begin{align}
	\mathcal{P}_{F}(x) & = \frac{1+(1-x)^{2}}{x}, &
	\mathcal{P}_{S}(x) & = \frac{2(1+x)}{x}.
\end{align}
We use these expressions to approximate the FSR contributions from two-body final states. For \(n\geq 3\) final state particles, we replace the
center-of-mass energy with the invariant mass of a pair of charged final states and marginalize over the invariant mass distributions; that is, we compute
\begin{align}
	\eval{\dv{N_{\rhn\to X + \gamma}}{x}}_{\mathrm{FSR}}
	 & \sim
	\frac{1}{2}\sum_{I,J\in X}\int \dd{M_{IJ}}
	P(M_{IJ})
	\qty[
	{\dv{N_{I^*\to I+\gamma}}{x}}\qty(x|M_{IJ}) +
	{\dv{N_{J^*\to J+\gamma}}{x}}\qty(x|M_{IJ})
	]\,,
\end{align}
where \(M_{IJ}=\sqrt{\qty(p_{I}+p_{J})^{2}}\) is the invariant mass of particles
\(I\) and \(J\), \(P(M_{IJ})\) is the distribution of invariant masses, and the factor of \(1/2\) accounts for double counting.

Recall that RHNs produced from dark matter decay or annihilation can be highly boosted if $m_\chi\gg m_N$. To generate the differential energy
spectrum \(\dv*{N_f}{E}\) from a boosted RHN, we first compute the spectrum in the RHN's rest frame, then boost it to the center of mass frame of dark matter annihilation or decay, which corresponds to the galactic frame, where these spectra are measured by experiments (see Appendix \ref{app:boost} for details of the boost procedure).

\subsection{Convolved Spectra}

To take into account the finite energy resolution of detectors and telescopes, we ``convolve'' the
spectra\footnote{
	More precisely, we marginalize over the conditional probability
	\(P(E|E')\) of an instrument detecting an energy \(E\) given the true energy
	\(E'\).
} with a spectral resolution function in order to provide more realistic spectra as would be observed by experiments. We adopt the standard approach of
modeling the energy resolution as a normal distribution with a standard
deviation proportional to the detector energy resolution \(\epsilon\)
\begin{align}
	R_{\epsilon}(E|E') = \frac{1}{\sqrt{2\pi}\epsilon E'}
	\exp(-\frac{1}{2}\qty(\frac{E-E'}{\epsilon E'})^{2}),
\end{align}
where \(E\) is particle energy measured by the detector and \(E'\) is the true
energy.  While the energy resolution is realistically a function of the
detected particle's energy, for simplicity, we use a constant energy resolution $\Delta E/E=0.05$
when displaying spectra. 

Taking into account a non-zero energy resolution, the differential energy
spectrum of a particle \(f\) is given by
\begin{align}
	\dv{N}{E}(E) = \int_{0}^{\infty}\dd{E'}R_{\epsilon}(E|E'){\dv{N}{E}}(E')\,.
\end{align}
The most important effect of including an energy resolution when displaying the spectra is the smearing of the photon, positron, and neutrino lines from processes such as \(\rhn \to e^{+}\pi^{-}, \gamma\nu\), and \(\nu\pi^{0}\): a monochromatic peak at \(\dv*{N}{E}\sim\delta\qty(E-E_0)\), where \(E_0\) is the energy of the stable product in the rest frame of the \(\rhn\), can only be resolved to \(\dv*{N}{E}\sim R_{\epsilon}\qty(E|E_0)\).

\section{Results}
\label{sec:results}

Using the above framework and formalism, we now present the main results of this paper: the decay branching ratios of RHNs, followed by the photon, positron, and neutrino spectra from DM annihilation or decay into RHNs over a broad range of mass scales.

\subsection{Sterile neutrino decay branching ratios}
\label{sec:BRs}

\begin{figure}
    \centering
    \includegraphics[width=\textwidth]{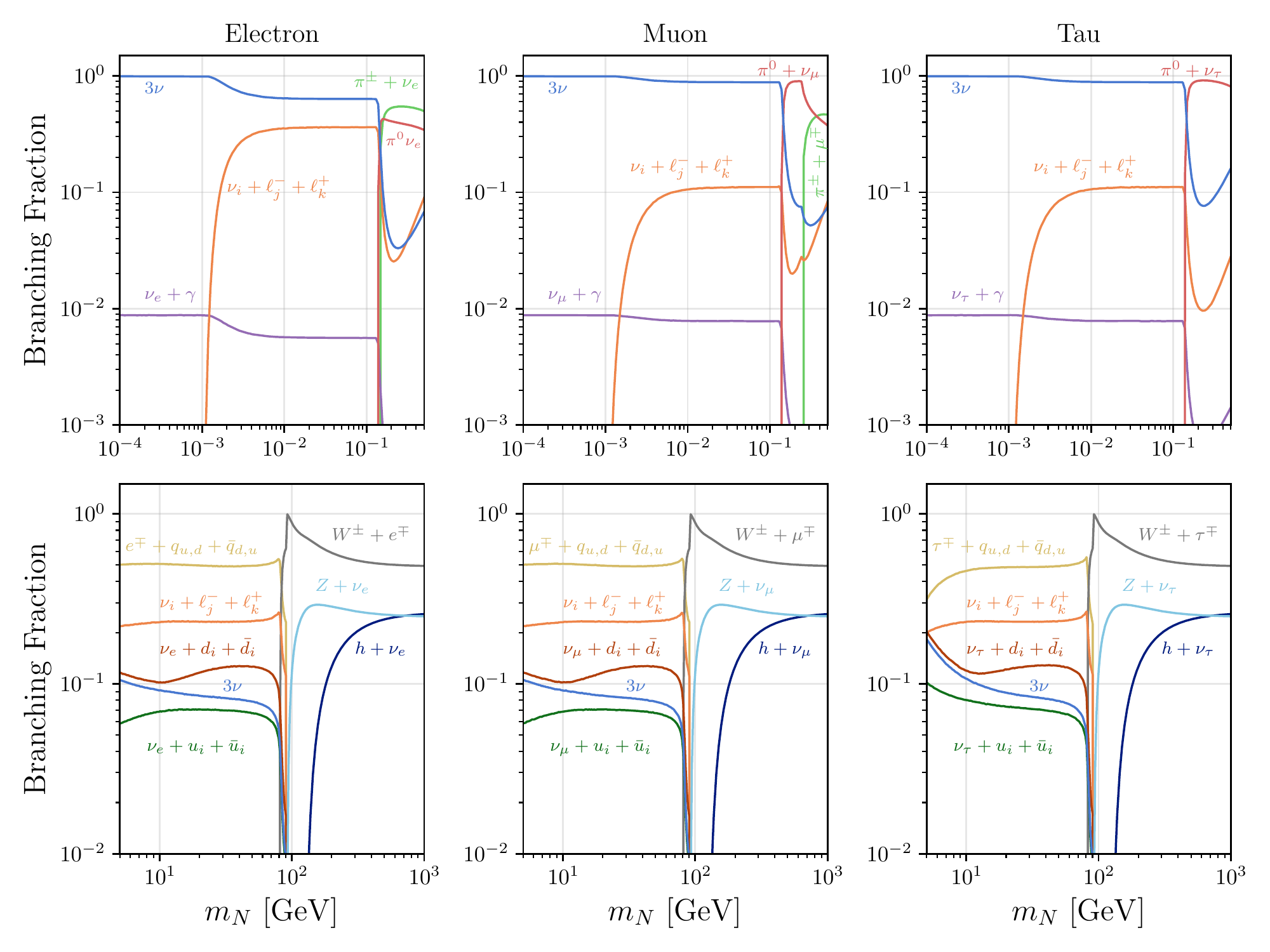}
    \caption{
        {{Upper (Lower) panels}}: Branching ratios for $m_{\rhn} < 500 \ \mathrm{MeV}$ ($m_{\rhn} > 5 \ \mathrm{GeV}$). 
        The RHN is assumed to mix with only one neutrino flavor; the three columns correspond to mixing with electron, muon, and tau neutrinos, respectively. Similar states have been grouped: e.g. \(3\nu\) and \(\nu+\ell_{i}^{\pm}+\ell_{j}^{\mp}\) include all possible combinations of neutrino and lepton flavors.
    }
    \label{fig:mrhn_mev}
\end{figure}

We first discuss the RHN branching ratios into various SM final states. 
These are presented in Fig.\,\ref{fig:mrhn_mev} for $m_{\rhn} < 500 \ \mathrm{MeV}$ in the upper panels, and $m_{\rhn} > 5 \ \mathrm{GeV}$ in the lower panels (recall that we do not study the $500 \ \mathrm{MeV}-5 \ \mathrm{GeV}$ regime due to complicated hadronic final states). For simplicity, we assume that the RHN mixes with a single flavor of SM neutrino; the three columns correspond to mixing with the electron, muon, or tau neutrino. 

In the light mass ($m_{\rhn} < 500 \ \mathrm{MeV}$) regime (upper panels), the main decay mode is the 3-neutrino channel (dark blue curves), until the 2-body decay into a lepton+pion final state opens up kinematically, which has an $\mathcal{O}(1)$ larger branching ratio due to phase space factors. The branching ratio into the $\nu e^+e^-$ channel (orange curve) is also comparable to the $3\nu$ channel above the $e^+e^-$ threshold (1 MeV) for the case of mixing with the electron neutrino, but is suppressed by about an order of magnitude for mixing with the muon and tau neutrinos, since for these latter cases the $\nu e^+e^-$ final state can only be produced by $Z$-mediated (instead of $W$-mediated) processes. However, for the case of mixing with the muon neutrino, note that the $\nu\mu^+\mu^-$ channel opens up close to 500 MeV, hence the total $\nu l^+ l^-$ branching ratio rises up to the same level as the $3\nu$ channel. On the other hand, the branching ratio into the $\nu\gamma$ final state is always less than percent level since this channel is loop-suppressed.

In the opposite regime of very large RHN masses (lower panels), where all lepton masses are negligible by comparison, the patterns are more flavor-universal. Above the electroweak gauge boson mass threshold, the dominant decay channel is $l^\pm W^\pm$ for all flavors, with $\nu_l  Z$ and $\nu_l h$ contributing between 10\% and 20\% of the decays. Below the $W$ mass threshold, the gauge bosons are off-shell, and the dominant channels are the $W$-mediated three body decays $l\,u\,d$ and $\nu l_i \bar l_j$, with contributions on the order of 10\% or lower from the $Z$-mediated $3\nu$, $\nu_l  d_i\bar d_i,$ and $\nu_l  u_i\bar u_i$ channels.

\subsection{Photon, positron, and neutrino spectra}

We now present the spectra of observable particles for various choices of $m_\chi$ and $m_N$. As mentioned earlier, we focus on the spectra of photons, positrons, and neutrinos, which could be detected by several current and upcoming experiments, but ignore antiprotons, for which the experimental reach is expected to be comparatively weaker in the models we consider.  

\vskip 0.2cm
\noindent\textit{Photons:} 

\begin{figure}
    \centering
    \includegraphics[width=\textwidth]{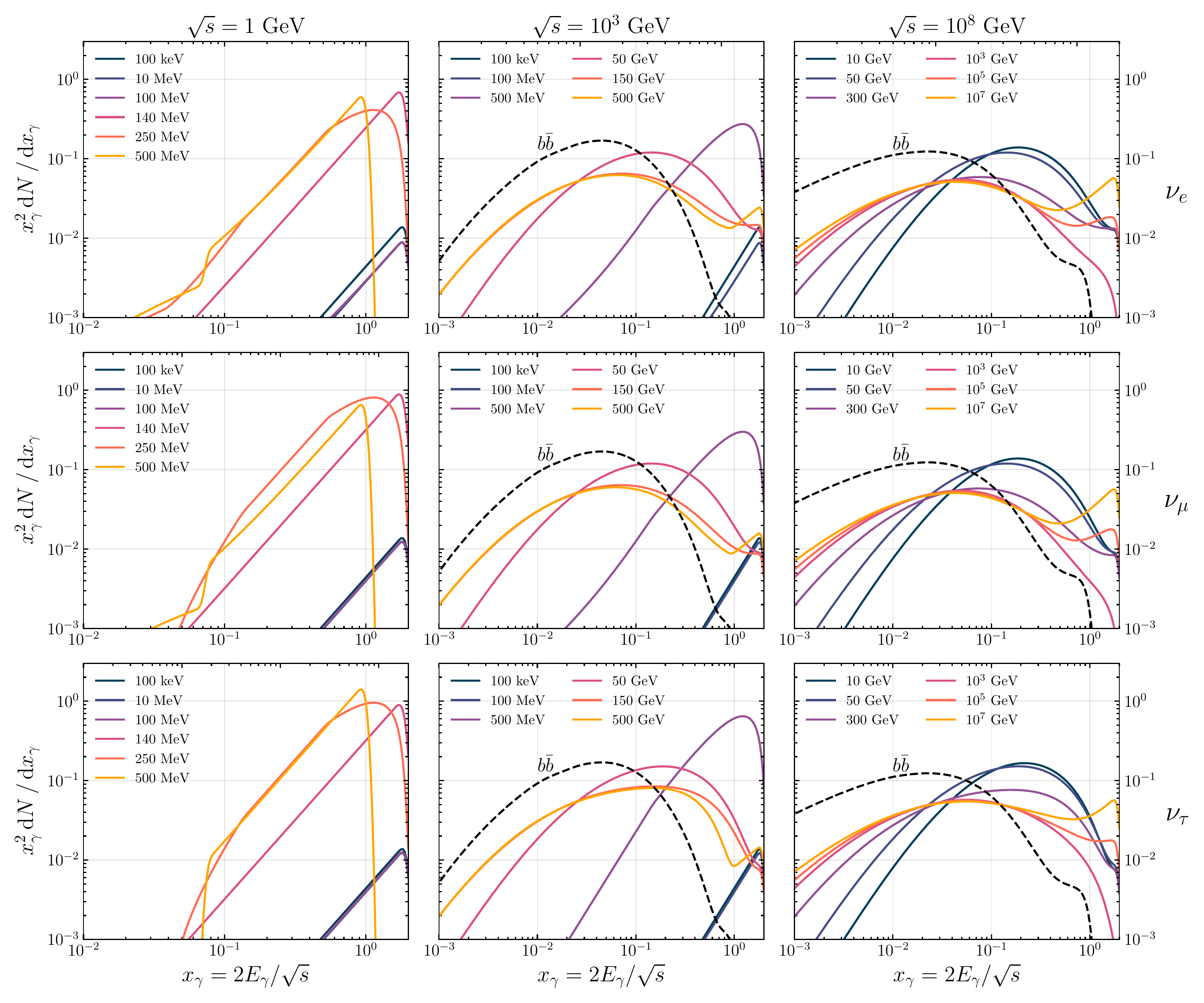}
    \caption{Photon spectra for dark matter annihilation or decay with center of mass energy $\sqrt{s}=1\ {\rm GeV}$ (left column), $\sqrt{s}=10^3\ {\rm GeV}$ (middle), and $\sqrt{s}=10^8\ {\rm GeV}$ (right), for mixing with the electron- (top row), muon- (center) and tau- (bottom) neutrinos, for a variety of RHN masses (colored curves, see legends). The black dashed curves in the middle and right columns correspond to the standard spectra from $DM\to b\bar b$ typically considered in the literature.} 
    \label{fig:dndx_photon_nine_panel}
\end{figure}

\begin{figure}
    \centering
    \includegraphics[width=\textwidth]{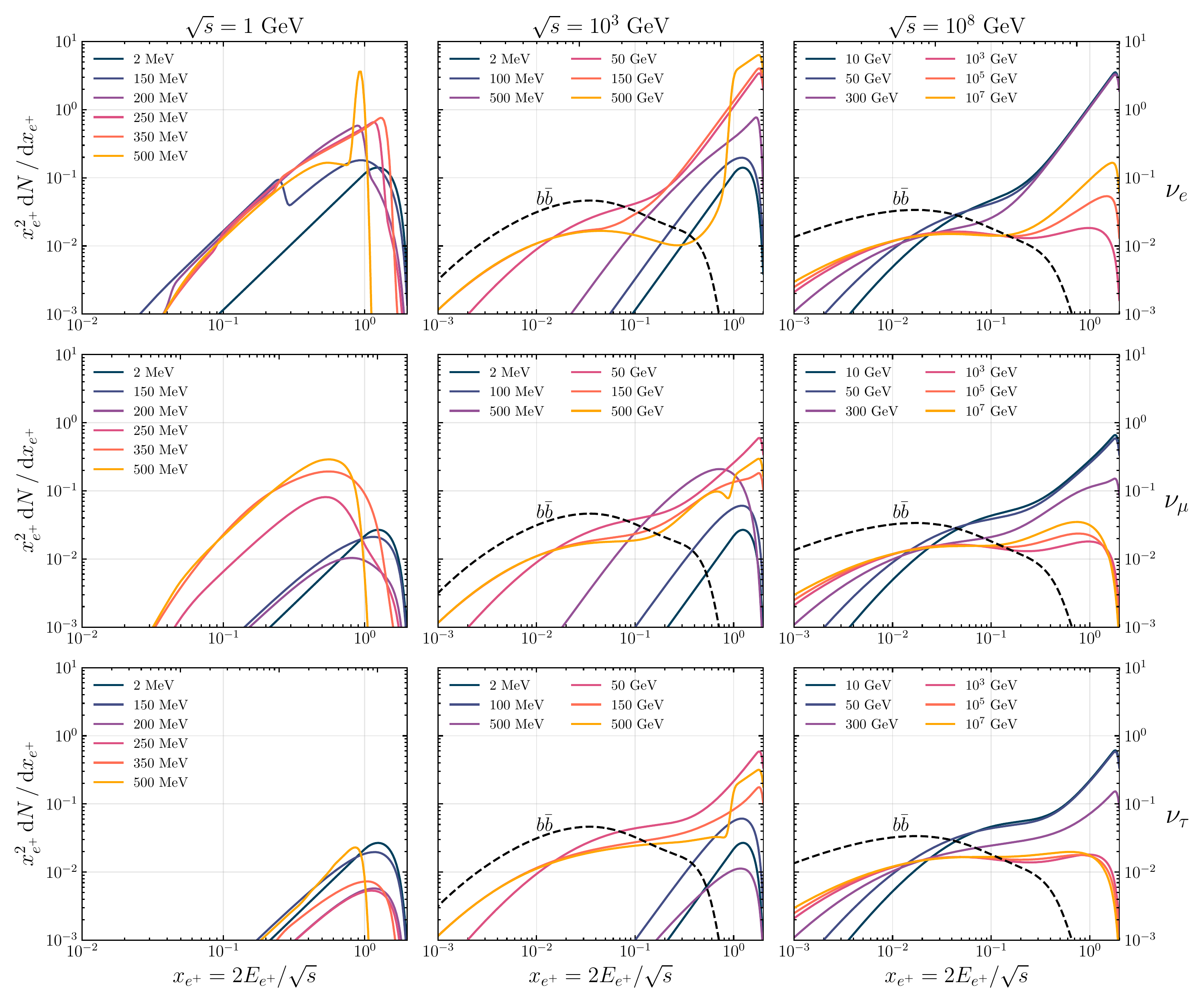}
    \caption{As in fig.~\ref{fig:dndx_photon_nine_panel}, but for the positron yield.
    }
    \label{fig:dndx_positron_nine_panel}
\end{figure}

\begin{figure}
    \centering
    \includegraphics[width=\textwidth]{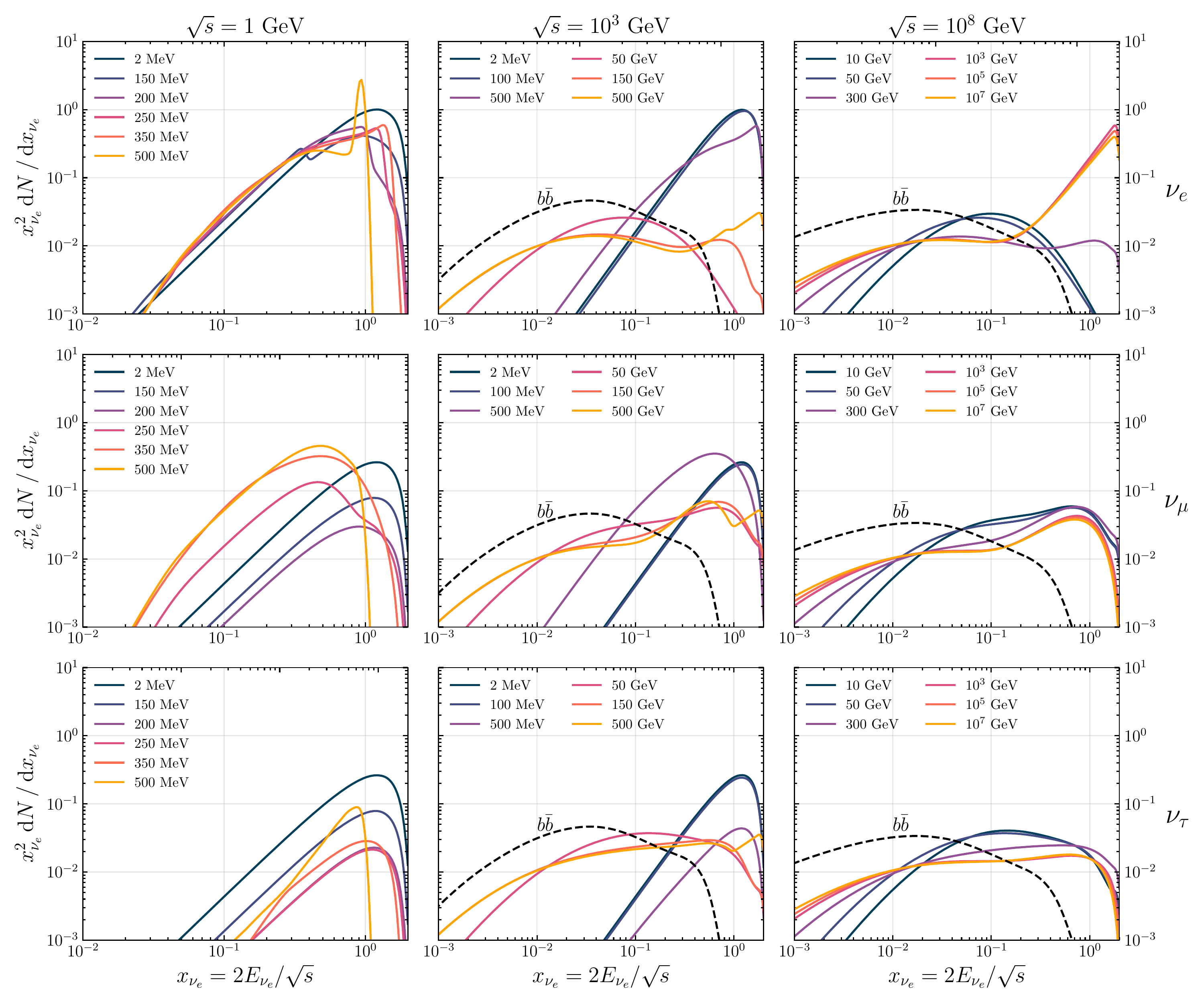}
    \caption{As in fig.~\ref{fig:dndx_photon_nine_panel}, but for the electron neutrino yield.
    }
    \label{fig:dndx_neutrino_nine_panel}
\end{figure}

In Fig.~\ref{fig:dndx_photon_nine_panel}, we present the computed spectra for photons for center of mass energies $\sqrt{s}=1,\ 10^3$ and $10^8$ GeV (left, center, and right columns, respectively); recall that $\sqrt{s}$ physically corresponds to $2m_\chi$ for dark matter annihilation, and to $m_\chi$ for decay; thus, the three panels represent GeV scale, TeV scale, and ultraheavy dark matter scenarios. The three rows correspond to the cases where the RHN mixes with the electron (top), muon (middle) and tau (bottom) neutrino, respectively. For each combination, we show the spectra for several choices of $m_N$, corresponding to the different colored curves in each panel (see plot legends for details). For comparison, we also show the photon spectra from dark matter directly decaying or annihilating into  $b\bar b$ (dashed black curves in the second and third columns), which is a standard benchmark for dark matter indirect detection studies. We use the same format to present the positron and neutrino spectra in Fig.~\ref{fig:dndx_positron_nine_panel} and Fig.~\ref{fig:dndx_neutrino_nine_panel}, respectively.

Photons are produced primarily in three ways: from the decays of neutral pions produced from decays of various SM states, from direct (loop) decays of the RHN, or as FSR from charged SM states. The former two contributions produce ``box-shaped" photon spectra $dN/dE\sim E^{0}$ if the SM decay products are highly boosted in the galactic frame (note that we plot $x^2 dN/dx$, so these spectra rise as $x^2$ instead of being box-shaped in our plots), whereas the FSR photon spectrum typically follows $dN/dE\sim E^{-1}$. In Fig.~\ref{fig:dndx_photon_nine_panel}, the loop decay $N\to\gamma \nu$ is prominent only for $m_N=100$ keV, since the only other available decay channel to $3\nu$ does not give rise to any photons; the box-shaped spectrum is clearly visible for this case. When the $\nu l^+l^-$ channel opens up, this adds a softer FSR tail to the above spectrum, as seen on the curves corresponding to $m_N=10, 100$ MeV (this feature is more prominent for mixing with the electron neutrino, as the branching ratio into charged leptons is higher, see Fig.\,\ref{fig:mrhn_mev}). For $m_{\pi}<m_N<1$ GeV, the dominant RHN decay channel is $\nu \pi^0$, leading to the aforementioned box-shaped photon spectrum again. For $5~\text{GeV}<m_N<m_W$, the dominant channel is $l+u+d$, and we see that the spectra are no longer box-shaped, but more closely resemble the canonical signals from $b\bar{b}$ (see curves for $m_N=10, 50$ GeV). For $m_N>m_W$, photons primarily arise from decays of boosted neutral pions produced by the fragmentation of gauge and Higgs bosons. As these pions are produced after multiple cascade steps, we see that the photon spectra are generally softer than those from cases where the RHN can decay directly to pions. In addition, these spectra feature rising peaks at high $x$ from FSR from the highly boosted $W$ bosons, which beomce more prominent for higher $m_N$. Also note that the spectra are essentially identical for mixing with electron and muon neutrinos (top two rows), whereas the case of mixing with the tau neutrino (bottom panel) gives rise to harder spectra due to a greater number of $\tau$ leptons in the decay products, which can produce pions more efficiently.

Overall, our results show that the photon spectra from DM annihilation/decay to RHNs are significantly harder than from those into $b\bar{b}$. Furthermore, we find that the spectral shape of the photon signal is primarily determined by $m_N$ (which determines the primary decay channels of the RHNs) and remains largely unaffected by the energy scale of the process $\sqrt{s}$. Therefore, if a photon signal is discovered, it is feasible that both the DM mass and the RHN mass can be inferred, from the energy scale and spectral shape of the signal respectively. On the other hand, determining which flavor of SM neutrino the RHN primarily couples to based on the observation of a photon signal appears to be challenging. 


\vskip 0.2cm
\noindent\textit{Positrons:} 

In Fig.~\ref{fig:dndx_positron_nine_panel}, we show our computed spectra for positrons; the format is identical to Fig.~\ref{fig:dndx_photon_nine_panel} except for slightly modified choices for $m_N$. It is crucial to note that these are positron spectra at production; we have not taken into account propagation effects between the point of production and the detector, which are known to be crucial for positrons in the intergalactic medium (see e.g. \cite{Moskalenko:1999sb,Delahaye:2007fr,Perelstein:2010fq,cirelli2011pppc}), and must be taken into account to derive spectra expected at detectors.

As was the case for the photon spectra, the most crucial parameter is $m_N$, which determines the dominant decay channels of the RHN. For the $m_N=2$ MeV curves in Fig.~\ref{fig:dndx_positron_nine_panel}, $N\to\nu e^+e^-$ is the only decay channel that contributes positrons, which is seen to give rise to a box-shaped spectra since the positrons are highly boosted. Note that the positron count is larger for mixing with the electron-type neutrino compared to the other two cases, as both $W$- and $Z$-mediated processes can produce this final state in the former case. Above the pion threshold, the $e^\pm\pi^\pm$ channel becomes the dominant source of positrons for mixing with $\nu_e$, hence all curves in the top left panel are very similar. For $m_N=500$ MeV, the RHN are effectively produced at rest, hence the decay $N\to e^\pm\pi^\pm$ produces a monochromatic positron line, which is clearly visible in this plot. On the other hand, for mixing with $\nu_\mu$ and $\nu_\tau$, the positron signals gets progressively weaker as the neutral decay channel $\nu_{\mu,\tau}\pi^0$ dominates, suppressing the $N\to\nu e^+e^-$ branching ratio; this trend is clearly visible for the $\nu_\tau$ mixing case (bottom left panel). For mixing with $\nu_\mu$, this is compensated for by the opening of the charged counterpart $N\to \mu^\pm\pi^\pm$, which can efficiently produce positrons; thus, in this case (middle left panel) the positron count goes down for $m_N=200, 250$ MeV, but rises again for $m_N=350, 500$ MeV. 

Above $m_N=5$ GeV, the crucial question is whether the $l^\pm W^\mp$ channel is kinematically open. Below this threshold, the main decay channel is $l u\bar{d}$, whereas above this threshold the main decay is into $l^\pm W^\mp$, where $l$ is of primarily the same flavor as the neutrino that $N$ mixes with. For mixing with $\nu_e$, both cases lead to similarly hard spectra at high $x$, as the emitted positron recoils against much heavier SM states, but the latter case leads to higher multiplicity of positrons at lower $x$ from the cascades of SM bosons. For mixing with $\nu_\mu$ and $\nu_\tau$, positrons are only produced from subsequent decays of the primary decay products, hence the positron spectrum declines far more rapidly at higher $x$ compared to the corresponding spectra with $\nu_e$ mixing. 
As was the case with the photon spectra, the qualitative features of the positron spectra also remain similar for higher energy scales $\sqrt{s}=10^3, 10^8$ GeV. In addition, for all of the plotted curves we see that the positron spectra are uniformly harder than those from $b\bar{b}$, although the positron count can be much lower in some cases. 


\vskip 0.2cm
\noindent\textit{Neutrinos:} 

In Fig.~\ref{fig:dndx_neutrino_nine_panel} we show similar spectra for neutrinos. Here we only show the spectra for $\nu_e$ as, for each row (representing mixing with a specific neutrino flavor), the spectra for $\nu_\mu, \nu_\tau$ can be inferred by permutation with the other two rows.   

From comparing with Fig.~\ref{fig:dndx_positron_nine_panel}, it is clear that the neutrino spectra follow essentially the same qualitative patterns as the positron spectra, as can be understood from the fact that the left-handed neutral and charged lepton interactions are related by $SU(2)$ symmetry. This is particularly true for cases with $m_N<500$ MeV, where the neutrino spectra are identical to those for positrons up to $\mathcal{O}(1)$ factors. At higher masses, however, there are a few notable differences. For $m_N=10,\,50$ GeV, the neutrinos primarily come from the $\nu l^+l^-$ and $3\nu$ decay channels, for which the spectra are softer and hence do not rise as sharply at large $x$ (as the positron spectrum does), as can be seen from comparing the corresponding curves in the second and third columns of the two figures. For $m_N\gg 100$ GeV, the main decay channel is $W^\pm+l^\mp$, followed by $Z+nu$; if $m_{DM}\gg m_N$, this results in highly boosted neutrinos that, again, follow a box-shaped spectrum, as seen clearly in the top right panel. 

Overall, we see the same general pattern as with photons and positrons: the spectra are uniformly harder than those from $b\bar{b}$, with lower counts.

\section{Summary}\label{sec:conclusions}

In this paper, we have provided a comprehensive study of indirect detection signals of dark matter annihilation or decay into right-handed, i.e. ``sterile'', neutrinos (RHN), which are characteristic of models where dark matter resides in a hidden sector that communicates with the visible sector via a neutrino portal. While previous papers only focused on specific, narrow mass ranges, we have provided a broad study that spans dark matter and RHN masses from $10^{-5}$ to $10^8$ GeV, which allows us to discuss indirect detection signals over a vast parameter space. 


The key difficulties in performing the analysis presented here for very low (below a few GeV) or very large (several TeV) RHN masses stem from the inability of Monte Carlo codes such as \Pythia\ and \Herwig\ to correctly compute the resulting decay products. 
We implemented the correct calculations in the low mass regime using the \Hazma\ tool, co-developed by two of the authors of this paper, and in the very large mass regime 
using the state-of-the-art code \HDMSpectra.

We presented the spectra of gamma rays, positrons, and neutrinos from dark matter annihilation or decay into sterile neutrinos for a broad range of dark matter and RHN masses, discussing important qualitative features of the spectra in various regimes.  Our results indicate that the spectral shape is primarily controlled by $m_N$; in particular, whether its mass is below the pion threshold (in which case decays into 3-body lepton states are dominant) or slightly above (decays into lepton+pion states dominate), and whether its mass is above or below the electroweak gauge boson mass scale (such that decays into lepton+gauge boson or 3-body fermion states are dominant, respectively). Thus, it is plausible that the discovery of a signal could provide information about dark matter mass from the overall energy scale, as well as the sterile neutrino mass from spectral features.

We also compared our spectra with generic expectations from DM decay/annihilation into SM final states widely considered in the literature ($b\bar b$), highlighting important differences. We found that the former spectra generally feature harder photon and neutrino emission at the highest possible energies than those from annihilation into $b\bar{b}$, thus leading to overall harder spectra, but with overall lower counts. 
These results suggest that the reach for dark matter annihilation or decay into RHNs using photon, positron, or neutrino observations should be comparable to the reach for DM annihilation/decay into $b\bar{b}$, with particularly promising prospects with gamma ray experiments such as HESS \cite{HESS:2022ygk}, Cherenkov Telescope Array (CTA) \cite{CTAConsortium:2017dvg}, and the Large High Altitude Air Shower Observatory
(LHAASO) \cite{DiSciascio:2015exu},  and neutrino telescopes such as IceCube \cite{IceCube:2022clp}. A detailed study of the reach of various current and future experiments for such dark matter signals is beyond the scope of this paper, but would be worthy of future study.

The code used in this study, and utilized to produce all figures shown here, is publicly available on \GitHubLink{GitHub}~\cite{BlackthornGitHub}.

\acknowledgments

We thank Stefania Gori for collaboration in the early stages of the project. LM and SP are supported in part by DOE grant DE-SC0010107. The work of BS is supported by the Deutsche Forschungsgemeinschaft under Germany‚ Excellence Strategy - EXC 2121 Quantum Universe - 390833306. BS thanks the Berkeley Center for Theoretical Physics and the Lawrence Berkeley National Laboratory for hospitality during the completion of this project. 

\appendix

\section{Matching onto Chiral Lagrangian}
\label{app:chiral}

In this appendix, we briefly discuss how to match the RHN Lagragian onto the chiral
Lagragian to obtain interactions between the RHNs and mesons; a more detailed discussion can be found in Ref.~\cite{Coogan:2021sjs}.

The leading-order chiral Lagrangian reads
\begin{align}
	\cL_{\ChiPT}
	 & =
	\frac{\fpi^{2}}{4}\Tr[\qty(D_{\mu}\bm{\Sigma})^{\dagger}\qty(D^{\mu}\bm{\Sigma})]
	+ \frac{\fpi^{2}}{4}\Tr[\bm{\Sigma}^{\dagger}\bm{\chi} + \bm{\chi}^{\dagger}\bm{\Sigma}],
\end{align}
where \(\fpi \sim 93\) MeV is the pion decay constant, \(\bm{\Sigma}\) is the
exponential of the Psuedo-Goldstone matrix, \(D_{\mu}\) is the chiral covariant derivative, and \(\bm{\chi}\) is a Spurion field mediating the soft chiral symmetry breaking. Explicitly, \(\bm{\Sigma}\) is given by:
\begin{align}
	\bm{\Sigma}              & = \exp(\frac{i}{\fpi}\phi^{a}\bm{\lambda}_{a}), &
	\phi^{a}\bm{\lambda}_{a} & =
	\mqty(
	\piz + \eta_{8}/\sqrt{3} & \sqrt{2}\pip                                    & \sqrt{2}\kp                 \\
	\sqrt{2}\pim             & -\piz +\eta_{8}/\sqrt{3}                        & \sqrt{2}\kz                 \\
	\sqrt{2}\km              & \sqrt{2}\kzb                                    & -\sqrt{\frac{2}{3}}\eta_{8} \\
	).
\end{align}
The chiral covariant derivative is given by
\begin{align}
	D_{\mu}\bm{\Sigma} & = \partial_{\mu}\bm{\Sigma} -i \bm{R}_{\mu}\bm{\Sigma} + i \bm{\Sigma}\bm{L}_{\mu},
\end{align}
with \(\bm{R}_{\mu}\) and \(\bm{L}_{\mu}\) representing the right and left
handed quark currents. The Spurion field \(\bm{\chi}\) is taken to have a
vacuum expectation value that breaks chiral symmetry in the same
way that quark masses do:
\begin{align}
	\bm{\chi} & = 2B_{0}\qty(\bm{M}_{q} + \bm{s} + i\bm{p}).
\end{align}
Here, \(B_{0}\) is related to the quark condensate via \(B_{0} \sim
-\expval{\bar{\bm{q}}\bm{q} + \bar{\bm{q}}^{\dagger}\bm{q}^{\dagger}} /
(3\fpi^{2})\) 
\footnote{The \(B_{0}\) parameter is ultimately removed in favor of the meson masses. For example, we can use \(m_{\pi^{\pm}}^{2} =B_{0}(m_{u}+m_{d})\)}
, and \(\bm{s}\) and \(\bm{p}\) are the scalar and
psuedo-scalar quark currents. The currents \(\bm{L}_{\mu}, \bm{R}_{\mu},
\bm{s}\) and \(\bm{p}\) are determined from the light-quark Lagrangian, written
as:
\begin{align}
	\cL =
	\cdots
	+ \bm{q}^{\dagger}\bm{L}^{\mu}\sibar_{\mu}\bm{q}
	+ \bar{\bm{q}}^{\dagger}\bm{R}^{\mu}\sibar_{\mu}\bar{\bm{q}}
	+ \bm{q}\qty(\bm{s}-i\bm{p})\bar{\bm{q}}
	+ \bm{q}^{\dagger}\qty(\bm{s}+i\bm{p})\bar{\bm{q}}^{\dagger},
\end{align}
where \(\bm{q}\) and \(\bar{\bm{q}}\) are the vectors containing the left and
right handed light quarks:
\begin{align}
	\bm{q}       & = \mqty(u       \\ d \\ s), &
	\bar{\bm{q}} & = \mqty(\bar{u} \\ \bar{d} \\ \bar{s}) .
\end{align}
Thus, matching onto the chiral Lagrangian simply requires us to identify the
quark currents. To do so, we begin by integrating out the heavy bosons. The result is the well-known
4-Fermi Lagrangian, given by:
\begin{align}\label{eqn:framework:four_fermi}
	\cL_{4F} & = -\frac{4\gf}{\sqrt{2}}\qty[J^{+}_{\mu}J^{-}_{\mu} + J^{Z}_{\mu}J^{Z}_{\mu}],
\end{align}
where the charged and neutral currents are given by:
\begin{align}
	J^{+}_{\mu}     & = \hat{\nu}_{i}^{\dagger}\sibar_{\mu}\ell_{i} + V_{ud}u^{\dagger}\sibar_{\mu}d + V_{us}u^{\dagger}\sibar_{\mu}s   ;      \\
	J^{-}_{\mu}     & = \ell_{i}^{\dagger}\sibar_{\mu}\hat{\nu}_{i} + V^{*}_{ud}d^{\dagger}\sibar_{\mu}u + V^{*}_{us}s^{\dagger}\sibar_{\mu}u;\\
	\cw J^{Z}_{\mu} & =
	g_{L,\nu}\hat{\nu}_{i}^{\dagger}\sibar_{\mu}\hat{\nu}_{i}
	+ g_{L,\ell}\ell_{i}^{\dagger}\sibar_{\mu}\ell_{i}
	+ g_{R,\ell}\bar{\ell}_{i}^{\dagger}\sibar_{\mu}\bar{\ell}_{i}
	+ \sum_{q=u,d,s}\qty(
	g_{L,q}q^{\dagger}\bar{\sigma}_{\mu}q + g_{R,q}\bar{q}^{\dagger}\bar{\sigma}_{\mu}\bar{q}),
\end{align}
where the sum over \(i\) is implied, \(V_{ud}\) and \(V_{us}\) are the \(u\)-\(d\)
and \(u\)-\(s\) CKM matrix elements, and the left/right-handed couplings are given by:
\begin{align}
	g_{L,u}    & = \frac{1}{2} - \frac{2}{3}\sw^2,  &
	g_{L,d}    & = -\frac{1}{2} + \frac{1}{3}\sw^2, &
	g_{L,\nu}  & = \frac{1}{2},                     &
	g_{L,\ell} & = -\frac{1}{2} + \sw^2,              \\
	g_{R,u}    & = -\frac{2}{3}\sw^2,               &
	g_{R,d}    & = \frac{1}{3}\sw^2,                &
	g_{R,\ell} & = \sw^2 .
\end{align}
Expanding out Eqn.\,(\ref{eqn:framework:four_fermi}), one can bring it into the following
form:
\begin{align}
	-\frac{\sqrt{2}}{4\gf}\cL_{4F}
	 & =
	\bm{q}^{\dagger}\qty[\bm{V}j_{\mu}^{-} + \bm{V}^{\dagger}j_{\mu}^{+} + \frac{2}{\cw}\bm{G}_{L,q}j_{\mu}^{Z}]\sibar_{\mu}\bm{q}
	+ \frac{2}{\cw}\bar{\bm{q}}^{\dagger}\qty(\bm{G}_{R,q}j_{\mu}^{Z})\sibar_{\mu}\bar{\bm{q}} \\
	 & \quad
	+ j_{\mu}^{+}j_{\mu}^{-} + j_{\mu}^{Z}j_{\mu}^{Z}
	+ \cdots \notag
\end{align}
where the \(\cdots\) contain terms that contribute to the next to leading order chiral Lagrangian, and \(j^{\pm}_{\mu}\) and \(j^{Z}_{\mu}\) are the charged and neutral currents containing leptons only:
\begin{align}
	j^{+}_{\mu}
	                                             & =
	\hat{\nu}_{i}^{\dagger}\sibar_{\mu}\ell_{i}, &
	j^{-}_{\mu}
	                                             & =
	\ell_{i}^{\dagger}\sibar_{\mu}\hat{\nu}_{i}      \\
	c_{W}j^{Z}_{\mu}
	                                             & =
	g_{L,\nu}\hat{\nu}_{i}^{\dagger}\sibar_{\mu}\hat{\nu}_{i}
	+ g_{L,\ell}\ell_{i}^{\dagger}\sibar_{\mu}\ell_{i}
	+ g_{R,\ell}\bar{\ell}_{i}^{\dagger}\sibar_{\mu}\bar{\ell}_{i}
\end{align}
The coupling matrices \(\bm{V}\), \(\bm{G}_{L,q}\) and \(\bm{G}_{R,q}\)
are given by:
\begin{align}
	\bm{V}
	        & =
	\mqty(0 & V_{ud} & V_{us} \\ 0&0&0\\ 0&0&0),
	        &
	\bm{G}_{L,q}
	        & =
	\bm{T}_{3,q} - s_{W}^{2}\bm{Q}_{q},
	        &
	\bm{G}_{R,q}
	        & =
	-s_{W}^{2}\bm{Q}_{q}
\end{align}
with \(\bm{Q}_{q}\) the light-quark charge matrix and \(\bm{T}_{3,q}\) a matrix
containing the light-quark weak-isospin eigenvalues
\begin{align}
	\bm{Q}_{q}
	 & =
	\frac{1}{3}\mqty(\dmat{2,-1,-1}),
	 &
	\bm{T}_{3,q}
	 & =
	\frac{1}{2}\mqty(\dmat{1,-1,-1}),
\end{align}
Written in this form, we can immediately identify the left and right handed currents as:
\begin{align}
	\bm{L}_{\mu}
	 & =
	\frac{2(1-s_{W}^{2})}{c_{W}}\qty(\bm{\lambda}_{3}+\frac{1}{\sqrt{3}}\bm{\lambda}_{8})j^{Z}_{\mu}
	+
	\qty[
	\qty(
	V^{*}_{ud}\bm{\lambda}^{+}_{12} + V^{*}_{us}\bm{\lambda}^{+}_{45}
	)j^{+}_{\mu}
	+\mathrm{c.c.}
	]\notag; \\
	\bm{R}_{\mu}
	 & =
	-\frac{2s_{W}^{2}}{c_{W}^{2}}
	\qty(\bm{\lambda}_{3}+\frac{1}{\sqrt{3}}\bm{\lambda}_{8})j^{Z}_{\mu},
\end{align}
where \(\bm{\lambda}^{\pm}_{12} = (\bm{\lambda}_{1} \mp i\bm{\lambda}_{2})/2\)
and \(\bm{\lambda}^{\pm}_{45} = (\bm{\lambda}_{4} \mp i\bm{\lambda}_{5})/2\).
Note that we have not included the singlet term of \(\bm{L}_{\mu}\) since the
vector singlet vanishes from the chiral covariant derivative and the axial
singlet must be taken into account via the chiral anomaly.

Expanding the chiral Lagrangian and dropping terms of order \(\order{G_{F}^{2}}\), we obtain:
\begin{align}
	\cL & =
	-\frac{4G_{F}}{\sqrt{2}}\qty(
	j^{Z}_{\mu}\cJ^{0}_{\mu}
	+ \qty(j^{+}_{\mu}\cJ^{+}_{\mu} + \mathrm{c.c.})
	+\qty(j^{Z}_{\mu})^{2}\cS^{0}
	+j^{Z}_{\mu}\qty(j^{+}_{\mu}\cS^{+}+\mathrm{c.c.})
	).
\end{align}
In the above expression, we organized the mesonic terms into charged and neutral vector and scalar currents/densities. The neutral meson current \(\cJ^{0}_{\mu}\) and the neutral density \(\cS^{0}\) are given by:
\begin{align}
	\cJ^{0}_{\mu} & =
	\frac{f_{\pi}}{c_{W}}\qty[
		\partial^{\mu}\pi^{0}
		+ \frac{1}{\sqrt{3}}
		\partial^{\mu}\eta
	]
	- \frac{i (1-2s_{W}^{2})}{c_{W}}
	\qty[
	\pi^{+}\overleftrightarrow{\partial}_{\mu}\pi^{-} +
	K^{+}\overleftrightarrow{\partial}_{\mu}K^{-}
	],\\
	\cS^{0} & =
	- \frac{4(1-s_{W}^{2})s_{W}^{2}}{c_{W}^{2}}
	\qty[
	\pi^{+}\pi^{-} +
	K^{+}K^{-}
	]
\end{align}
where \(X\overleftrightarrow{\partial}_{\mu}Y \equiv X\partial_{\mu}Y - Y\partial_{\mu}X\).
The charged meson current and density are:
\begin{align}
	\cJ^{+}_{\mu}
	 & =
	\frac{f_{\pi}}{\sqrt{2}}\qty[
	V^{*}_{ud}\partial^{\mu}\pi^{+}
	+ V^{*}_{us}\partial^{\mu}K^{+}
	]
	\\
	 & \quad
	+ i
	\frac{V^{*}_{ud}}{2}\qty(
	\bar{K}^{0}\overleftrightarrow{\partial}_{\mu}K^{+}
	+ \sqrt{2}\pi^{+}\overleftrightarrow{\partial}_{\mu}\pi^{0}
	)\notag  \\
	 & \quad
	+i\frac{V_{us}^{*}}{2}\qty(
	K^{0}\overleftrightarrow{\partial}_{\mu}\pi^{+}
	+\sqrt{\frac{3}{2}}K^{+}\overleftrightarrow{\partial}_{\mu}\eta
	+\frac{1}{\sqrt{2}}K^{+}\overleftrightarrow{\partial}_{\mu}\pi^{0}
	).\notag\\
	\cS^{+}
	 & =
	\frac{i\sqrt{2}\fpi\sw^2}{\cw}\qty(\vudc\pip + \vusc\kp) \\
	 & \quad
	+ \frac{s_{W}^{2}}{c_{W}}\qty(
	V^{*}_{ud}\qty(
	\sqrt{2}\pi^{+}\pi^{0}
	- K^{+}\bar{K}^{0}
	)
	+V^{*}_{us}\qty(
	\sqrt{\frac{3}{2}}K^{+}\eta
	+ \frac{1}{\sqrt{2}}K^{+}\pi^{0}
	- \pi^{+}K^{0}
	)
	)\notag
\end{align}

\section{Boosted Spectra}\label{app:boost}

In this appendix, we describe the procedure to boost the decay spectra obtained in the RHN rest frame to the center of mass frame of dark matter annihilation or decay.

Without loss of generality, we take the RHN to be boosted in the \(z\)-direction. We orient the 4-momentum of the particle \(f\) in the
RHN rest-frame to be in the \(xz\)-plane
\begin{align}
	p_{1}^{\mu}=\qty(E_{1}; \bm{p}_{1}) = \qty(E_{1}; \sqrt{1-z^2}\abs{\bm{p}_{1}}, 0, z\abs{\bm{p}_{1}}).
\end{align}
Here, \(z=\cos\theta\) is the cosine of the angle with respect to the
\(z\)-axis, \(E_{1}\) is the energy of \(f\) in the RHN rest frame, and
\(\abs{\bm{p}_1} =\sqrt{E_{1}^2 - m_{f}^2}\). If the RHN has an energy
\(E_{\rhn}\) in the lab-frame, then the boosted four-momentum is
\begin{align}
	p_{2}^{\mu}=\qty(E_{2}; \bm{p}_{2}) = \qty(
	\gamma E_{1} + \beta\gamma\abs{\bm{p}} z,
	\sqrt{1-z^2}\abs{\bm{p}_{1}}, 0, \beta\gamma E_{1} + \gamma\abs{\bm{p}_{1}} z
	),
\end{align}
where \(\gamma = E_{\rhn}/m_{\rhn}\), and \(\beta=\sqrt{1-1/\gamma^{2}}\) is the
velocity of the RHN. Take the differential energy spectrum of the
RH-neutrino in its rest-frame to be \(\dv*{N_{f}}{E_1}\). We can reintroduce the
angular dependence by multiplying the spectrum by \(\frac{1}{2}\int_{-1}^{1}\dd{z}\).
To change variables to the lab-frame, we add an integral over a
\(\delta\)-function, which forces the correct relationship between the rest-frame
quantities and the lab-frame energy \(\delta(\gamma E_{1} + \beta\gamma\abs{\bm{p}_{1}} z - E_2)\).
After these modifications, the average differential number of \(f\)s produced
per decay/annihilation can be written as
\begin{align}
	\dd{N_{f}} = \dd{E_1}\dv{N_{f}}{E_1} = \frac{1}{2}\dd{E_2}\dd{E_1}\dd{z}\dv{N_{f}}{E_1}
	\delta(\gamma E_{1} + \beta\gamma\abs{\bm{p}_{1}} z - E_2).
\end{align}
Using the \(\delta\)-function, we can integrate over the angular variable \(z\), arriving at the resulting spectrum in the lab-frame
\begin{align}
	\dv{N_{f}}{E_2} = \frac{1}{2\gamma\beta}\int^{E^+_1}_{E^-_1}\frac{\dd{E_1}}{\sqrt{E_1^2-m_1^2}}{\dv{N_{f}}{E_1}}\qty(E_1).
\end{align}
The integration bounds are determined from the support of the introduced
\(\delta\)-function and the limits \(-1 \leq z \leq 1\). Explicitly, the
bounds are
\begin{align}
	E^{+}_{1} & = \mathrm{min}\qty(
	E_{1}^{\mathrm{max}},
	\gamma\qty(E_{2} + \beta\sqrt{E_{2}^{2} - m_{f}^{2}})
	),                              \\
	E^{-}_{1} & = \mathrm{max}\qty(
	E_{1}^{\mathrm{min}},
	\gamma\qty(E_{2} - \beta\sqrt{E_{2}^{2} - m_{f}^{2}})
	).
\end{align}
In these expressions, the quantities \(E^{\mathrm{min}}_{1}\) and
\(E^{\mathrm{max}}_{1}\) are the minimum and maximum energies of \(f\) in the original
frame, beyond which \(\dv*{N_{f}}{E_{1}}\) is zero. In terms of dimensionless energy
variables \(x_{1} = 2E_{1}/m_{\rhn}\) and \(x_{2} = 2E_{2}/E_{\rhn}\),
the spectrum is:
\begin{align}
	\dv{N_{f}}{x_{2}} & =
	\frac{1}{2\beta}
	\int^{x_{1}^{+}}_{x_{1}^{-}}
	\frac{\dd{x_{1}}}{\sqrt{x_{1}^{2} - \mu_{1}^{2}}}
	\dv{N_{f}}{x_{1}}
\end{align}
with \(\mu_{1} = 2m_{f}/Q_{1}\) and
\begin{align}
	x^{+}_{1} & = \mathrm{min}\qty(
	x_{1}^{\mathrm{max}},
	\gamma^{2}\qty(x_{2} + \beta\sqrt{x_{2}^{2} - \mu_{2}^{2}})
	),                              \\
	x^{-}_{1} & = \mathrm{max}\qty(
	x_{1}^{\mathrm{min}},
	\gamma^{2}\qty(x_{2} - \beta\sqrt{x_{2}^{2} - \mu_{2}^{2}})
	).
\end{align}

\bibliographystyle{JHEP}
\bibliography{DMRHN}



\end{document}